\documentclass{article}
\usepackage{amsmath,amssymb,bm,braket,a4wide,parskip,bbold,color,xcolor,subcaption,url,graphicx,hyperref,titling}

\usepackage[frozencache,cachedir=.]{minted}

\definecolor{mypink}{rgb}{1,.443,.830}
\hypersetup{
    colorlinks = true,
    linkcolor = mypink,
    citecolor = mypink,
    urlcolor = gray
}

\def\matr#1{\bm{#1}}
\def\iu{\mathrm{i}}
\def\xor{\oplus}

\title{
    Decomposing dense matrices into dense Pauli tensors
}
\author{Tyson Jones\thanks{tyson.jones.input@gmail.com}
\\ 
\footnotesize
Department of Materials, University of Oxford, Parks Road, Oxford OX1 3PH, United Kingdom
\\
\footnotesize
Institute of Physics, Ecole Polytechnique Fédérale de Lausanne (EPFL), CH-1015 Lausanne, Switzerland
\\ 
\footnotesize
Quantum Motion Technologies, Pearl House, 5 Market Road, London N7 9PL, United Kingdom
}

\date{January 2024}

\begin{document}

\maketitle

\begin{abstract}
    Decomposing a matrix into a weighted sum of Pauli strings is a common chore of the quantum computer scientist, whom is not easily discouraged by exponential scaling. But beware, a naive decomposition can be \textit{cubically} more expensive than necessary!
    In this manuscript, we derive a fixed-memory, branchless algorithm to compute the inner product between a $2^N\times 2^N$ complex matrix and an $N$-term Pauli tensor in $\mathcal{O}(2^N)$ time, by leveraging the Gray code. Our scheme permits the embarrassingly parallel decomposition of a matrix into a weighted sum of Pauli strings in $\mathcal{O}(8^N)$ time. We implement our algorithm in Python, hosted open-source on Github~\cite{tysongithub}, and benchmark against a recent state-of-the-art method called the ``PauliComposer"~\cite{romero2023paulicomposer,romerogithub} which has an exponentially growing memory overhead, achieving speedups in the range of $1.5\times$ to $5\times$ for $N \le 7$.
    Note that our scheme does not leverage sparsity, diagonality, Hermitivity or other properties of the input matrix which might otherwise enable optimised treatment in other methods. As such, our algorithm is well-suited to decomposition of {dense}, arbitrary, complex matrices which are expected dense in the Pauli basis, or for which the decomposed Pauli tensors are \textit{a priori} unknown.
\end{abstract}

\section{Introduction}

Given a dense complex matrix $\matr{G} : \mathbb{C}^{2^N \times 2^N}$, we seek $4^N$ coefficients $c_i \in \mathbb{C}$ which encode $\matr{G}$ as a $4^N$-term sum of weighted $N$-Pauli tensors;
\begin{align}
    \matr{G} 
    \equiv \sum\limits_n^{4^N} 
    c_n \, \matr{P}_n
\end{align}
where $\matr{P}_n : \mathbb{C}^{2^N \times 2^N}$ are the matrices of the $N$-dimensional Pauli tensors.
For example, given $\matr{G}:\mathbb{C}^{4\times4}$, we seek $c_i$ satisfying
\begin{align}
    \matr{G} \equiv
    c_0 \, \matr{\mathbb{1}} \otimes \matr{\mathbb{1}}
    \; + \;
    c_1 \, \matr{\mathbb{1}} \otimes \matr{X}
    \; + \;
    c_2 \, \matr{\mathbb{1}} \otimes \matr{Y}
    \; + \;
    c_3 \, \matr{\mathbb{1}} \otimes \matr{Z}
    \; + \;
    c_4 \, \matr{X} \otimes \matr{\mathbb{1}}
    \; + \;
    \dots
    \; + \;
    c_{15} \, \matr{Z} \otimes \matr{Z},
\end{align}
where the above are the standard Pauli matrices 
\begin{align}
    \matr{\mathbb{1}} = \begin{pmatrix} 1 \\ & 1 \end{pmatrix},
    \;\;\;
    \matr{X} = \begin{pmatrix} & 1 \\  1 \end{pmatrix},
    \;\;\;
    \matr{Y} = \begin{pmatrix} & -\iu \\ \iu \end{pmatrix},
    \;\;\; 
    \matr{Z} = \begin{pmatrix} 1 \\ & -1 \end{pmatrix}.
\end{align}
Obtaining $c_i$ is a common task of a quantum computer scientist, who seeks to decompose a Hermitian operator - like a Hamiltonian, or a unitary operator's generator - into a real-weighted sum of Pauli strings. This permits its expectation evaluation on a quantum computer~\cite{huang2021efficient}, its Trotterisation into Pauli gadgets~\cite{hatano2005finding}, or the sampling of variational quantities~\cite{yuan2019theory}.

While the coefficients satisfy
\begin{align}
    c_n = \frac{1}{2^N} \text{Tr} \left( \,
        \matr{P}_n \; \matr{G}
    \,
    \right),
    \label{eq:cn_compact}
\end{align}
this is an impractical form to evaluate directly, and might naively imply a cost of $\mathcal{O}(8^N)$ \textit{per} $c_n$, or a total decomposition cost of $\mathcal{O}(32^N)$. Even by making no assumptions about $\matr{G}$, we can find a significantly more efficient evaluation by studying $\matr{P}_n$ and leveraging the properties of the Pauli matrices. In the next section, we derive a scheme which makes use of the fact that the Pauli matrices are diagonal or anti-diagonal, and that the elements of $\matr{P}_n$ can be enumerated in a deliberate order so as to speedup evaluation of the trace.

%




\section{Derivation}

Because our scheme makes extensive use of indexing and bitwise algebra, we adopt an unconventional notation.
We notate the $(i,j)$-th element of a matrix $\matr{M}$ in square brackets as $[\matr{M}]_{ij}$, where $i$ and $j$ begin at \textit{zero}. The $t$-th bit (also indexing from zero) of a non-negative integer $i \in \mathbb{N}^0$ is notated as $i_{[t]}$. The zero-th bit is the \textit{rightmost}, as is the zero-th operator in a Pauli tensor. All sum notation implicitly begins at zero and ends \textit{exclusive}, so that $\sum_i^3 x_i = x_0 + x_1 + x_2$.

We begin by expanding Eq.~\ref{eq:cn_compact} to a sum of $2^{2N}$ terms;
\begin{align}
c_n 
    &= \frac{1}{2^N}
    \sum_{i}^{2^N} 
     \sum_{j}^{2^N} 
     \left[ \matr{P}_n \right]_{ij}
     \left[ \matr{G} \right]_{ji}.
     \label{eq:c_def_elements_expanded}
\end{align}
Let the $n$-th Pauli tensor $P_n$ have individual Pauli operators $\sigma_t^{(n)}$, with matrices $\matr{\sigma}_t^{(n)} : \mathbb{C}^{2\times 2}$,
such that
\begin{align}
    \matr{P}_n \equiv \bigotimes\limits_t^N \matr{\sigma}_t^{(n)},
    \;\;\;\;\;\;\;
    {\sigma}_t^{(n)} \, \in \{ \mathbb{1}, X, Y, Z\},
\end{align}
to make explicit that an element of $\matr{P}_n: \mathbb{C}^{2^N\times 2^N}$ is given by 
\begin{align}
\left[ 
    \matr{P}_n
\right]_{ij}
    = 
    \prod\limits_{t}^N
    \left[ \matr{\sigma}_t^{(n)} \right]_{i_{[t]}, \, j_{[t]}}.
    \label{eq:pauli_tensor_elems_bitwise}
\end{align}
Such a bitwise treatment of a Pauli tensor is demonstrated in Ref.~\cite{jones2023distributed}, but we will here further leverage that the Pauli matrices are diagonal or anti-diagonal. As such, we will somewhat unusually label their elements as
\begin{align}
    \left[ \matr{\sigma}_t^{(n)} \right]_{ab} = \beta_{ab}^{(n,t)} 
    \, \times 
    \begin{cases}
        \delta_{ab} & 
            \sigma_t^{(n)} \in \{\mathbb{1}, Z\}
            \\
        1 - \delta_{ab} &
            \sigma_t^{(n)} \in \{X, Y\}
    \end{cases},
    \;\;\;\;\;\;\;\;
    \beta_{ab}^{(n,t)} \in \{\pm 1, \pm \iu\},
\end{align}
where $\delta_{ab}$ is the Kronecker delta. 
By substituting these forms back into $c_n$, we obtain
\begin{align}
c_n &= 
    \frac{1}{2^N}
    \sum_{i}^{2^N} 
     \sum_{j}^{2^N} 
     \left[ \matr{G} \right]_{ji}
     \,
    \prod\limits_t^N \beta_{i_{[t]}, j_{[t]}}^{(n,t)} 
        \times
        \begin{cases}
        \delta_{i_{[t]}, j_{[t]}} & 
            \sigma_t^{(n)} \in \{\mathbb{1}, Z\}
            \\
        1 - \delta_{i_{[t]}, j_{[t]}} &
            \sigma_t^{(n)} \in \{X, Y\}
    \end{cases}
    \\
    &=
\frac{1}{2^N}
    \sum_{i}^{2^N} 
    \left[ \matr{G} \right]_{f_n(i), i}
    \prod\limits_t^N
    \beta_{i_{[t]}, f_n(i)_{[t]} }^{(n,t)},
\end{align}
where function $f_n(i)$ accepts the $N$-bit integer $i$ and flips the bits at indices $t$ satisfying $\sigma_t^{(n)} \in \{X, Y\}$. 
We will construct (and in our algorithm, \textit{evaluate}) function $f_n$ using an $N$-length bitmask $m^{(n)}$ which contains a $1$ at position $t$ only if the $n$-th Pauli tensor $P_n$ contains an $X$ or $Y$ operator at position $t$. Then
\begin{align}
    f_n(i) = i  \xor  m^{(n)},
    \hspace{2cm}
    \therefore 
    f_n(i)_{[t]} = i_{[t]}  \xor {m^{(n)}}_{[t]} \,.
\end{align}
These bitwise evaluations are constant time on systems with a fixed-size word, and when $N$ is smaller or equal to that word size. For example, using a $32$-bit natural number bounds $N \le 32$. Or, if we instantiated $m^{(n)}$ as a \texttt{unsigned long long int} in a \texttt{C} program, evaluating $f_n(i)$ is $\mathcal{O}(1)$ for $N \le 64$.

We finally define
\begin{align}
\lambda_i^{(n)} = 
    \prod\limits_t^N
    \beta_{i_{[t]}, \, i_{[t]}  \xor {m^{(n)}}_{[t]} }^{(n,t)}
    \; \;
    \in \{\pm 1, \pm \iu \}
\label{eq:lambda_def}
\end{align}
so that we may express
\begin{align}
c_n &=
\frac{1}{2^N}
    \sum_{i}^{2^N} 
    \lambda_i^{(n)}
    \left[ \matr{G} \right]_{i  \xor  m^{(n)}, \, i}.
\label{eq:c_def_in_terms_of_fn}
\end{align}
So far, we have simplified Eq.~\ref{eq:c_def_elements_expanded} to a sum of only $2^N$ terms; a quadratic improvement. Evaluation of Eq.~\ref{eq:c_def_in_terms_of_fn} for a given $n$ still suggests $2^N$ independent evaluations of $\lambda_i^{(n)}$, each a product of $N$ scalars. We can further eliminate a factor $N$ in runtime by replacing an ordered iteration of $i \in [0\, .. \, 2^N)$ with an enumeration of the $N$-bit Gray-codes~\cite{graycoderev}. Subsequent $i$, which can be calculated in a fixed number of operations~\cite{donald1999art}, then differ by a single bit (a Hamming distance of $1$), enabling a recurrence in the evaluation of $\lambda_i^{(n)}$ across $i$.
This is by observing that only a \textit{single} scalar $\beta^{(n,t)}_{ab}$ among the $N$-term product in Eq.~\ref{eq:lambda_def} has changed as $i \rightarrow i'$.
Explicitly, let $i'$ differ from $i$ by a single bit at position $t$. Then
\begin{align}
    \lambda_{i'}^{(n)} = 
        \lambda_{i}^{(n)}  \;
        \beta_{i'_{[t]}, \, i'_{[t]}  \xor {m^{(n)}}_{[t]}
    }^{(n,t)} \; / \;
        \beta_{i_{[t]}, \, i_{[t]}  \xor {m^{(n)}}_{[t]} 
    }^{(n,t)},
    \label{eq:lambda_recurrence_over_i}
\end{align}
and ergo $\lambda_{i'}^{(n)}$ can be calculated from $\lambda_{i}^{(n)}$ in a fixed number of operations.

If we wished, we could also leverage a potential recurrence in $\lambda_i^{(n)}$ across $n$. We could enumerate the Pauli tensors $P_n$ via the \textit{quaternary} Gray code~\cite{er1984generating} such that $P_{n'}$ differs from the previously enumerated $P_n$ by a single Pauli operator at position $t$. Then
\begin{gather}
    m^{(n')} = \begin{cases}
        m^{(n)}, &
            \mathbb{1} \leftrightarrow Z \text{ or } X \leftrightarrow Y
            \\
        m^{(n)} \xor (1 \texttt{<<} t)
            &
                \text{otherwise}
    \end{cases},
    \\
    \lambda_{i}^{(n')} = 
        \lambda_{i}^{(n)}  \;
        \beta_{i_{[t]}, \, i_{[t]}  \xor {m^{(n')}}_{[t]}
    }^{(n',t)} \; / \;
        \beta_{i_{[t]}, \, i_{[t]}  \xor {m^{(n)}}_{[t]} 
    }^{(n,t)}\,.
\end{gather}
In serial settings, this enables us to avoid the initial $\mathcal{O}(N)$ calculation of $\lambda_0^{(n)}$ for each $n \in [0\,..\,4^N)$. This is a modest gain however, and in parallel settings, we should instead exploit the otherwise embarrassingly parallel evaluation possible of $c_n$ across $n$.


Note that use of the above recurrences require we know the position $t$ of the single bit $i'$ which has differed from the previous Gray code $i$. For an unbounded $N$, this determination costs time $\mathcal{O}(\log N)$, but with bounded $N$ as enforced by our aforementioned use of a bitmask, the index $t$ can be determined in constant time using a lookup table~\cite{donald1999art}, or a fixed-word bitwise calculation of $\log_2(i \xor i')$~\cite{anderson2005bit}.

Also note that our formulation assumed we can efficiently identify which Pauli operator (of $\mathbb{1},X,Y,Z$) admits scalar  $\beta^{(n,t)}_{ab}$, given the Pauli tensor index $n \in [0\,..\,4^N)$ and Pauli position index $t \in [0\,..\,N)$. This is equivalent to finding the $t$-th digit of a base-$4$ $N$-digit numeral, and can only be done in constant time if we again assert boundedness of $N$. We merely group together contiguous pairs of bits of $n$, with the $t$-th group encoding the flag $\in [0\,..4)$ of $\sigma^{(n)}_t \in \{\mathbb{1},X,Y,Z\}$. Beware that this has \textit{halved} the maximum number of Pauli tensor terms $N$ to be \textit{half} the number of bits in our natural number type.

\section{Algorithm}

Our algorithm to determine $c_n$ is merely to evaluate Eq.~\ref{eq:c_def_in_terms_of_fn}, 
enumerating the Gray codes in lieu of contiguously iterating the sum index $i$, and to use the constant-time recurrent definition of $\lambda_i^{(n)}$ given in Eq.~\ref{eq:lambda_recurrence_over_i}.
A high-level Python implementation of this scheme is given in Fig.~\ref{fig:alg}, and provided on Github~\cite{tysongithub}.
We note that in our subsequent testing and benchmarking, we use an alternate implementation which simply unwraps each function in Fig.~\ref{fig:alg}~(a) to be in-line, which approximately halves the runtime.

\begin{figure*}[p]
\centering 
\begin{subfigure}[t]{0.49\textwidth}
\begin{minted}{python}
def getPowerOf2(n):
    return 1 << n

def getLog2(n):
    r  = (n & 0xAAAAAAAA)  != 0
    r |= ((n & 0xFFFF0000) != 0) << 4
    r |= ((n & 0xFF00FF00) != 0) << 3
    r |= ((n & 0xF0F0F0F0) != 0) << 2
    r |= ((n & 0xCCCCCCCC) != 0) << 1
    return r

def getBit(n, t):
    return (n >> t) & 1

def getGrayCode(n):
    return n ^ (n >> 1)

def getChangedBit(i, j):
    return getLog2(i ^ j)

def getPauliFlag(n, t):
    b0 = getBit(n, 2*t)
    b1 = getBit(n, 2*t+1)
    return (b1 << 1) | b0

def getMaskOfXY(n, N):
    m = 0
    for t in range(N):
        p = getPauliFlag(n, t)
        b = 0 < p < 3
        m |= b << t
    return m

def getBeta(n, t, i, m):
    paulis = [
        [[1,0],[0,1]],
        [[0,1],[1,0]],
        [[0,-1j],[1j,0]],
        [[1,0],[0,-1]]]
    p = getPauliFlag(n, t)
    a = getBit(i, t)
    b = a ^ getBit(m, t)
    e = paulis[p][a][b]
    return e

def getLambda(n, i, m, N):
    l = 1
    for t in range(N):
        l *= getBeta(n, t, i, m)
    return l
\end{minted}
\caption{Convenience functions used by our algorithm.}
\end{subfigure} %
\begin{subfigure}[t]{0.49\textwidth}
\begin{minted}{python}
def calcPauliCoeffFast(n, G, N):
    c = 0
    m = getMaskOfXY(n, N)
    dim = getPowerOf2(N)

    # evaluate lambda_0 in full
    l = getLambda(n, 0, m, N)
    c += l * matrix[m][0]

    # recurrently evaluate lambda_i
    for k in range(1, dim):
        i = getGrayCode(k)
        j = getGrayCode(k - 1)
        t = getChangedBit(i, j)
        bi = getBeta(n, t, i, m)
        bj = getBeta(n, t, j, m)
        f = i ^ m

        # by revising one beta
        l *= bi / bj
        c += l * G[f][i]

    c /= float(dim)
    return c

def calcPauliCoeffSlow(n, G, N):
    c = 0
    m = getMaskOfXY(n, N)
    dim = getPowerOf2(N)

    # evaluate each lambda_i independently
    for i in range(dim):
        l = getLambda(n, i, m, N)
        c += l * G[i ^ m][i]
    
    c /= float(dim)
    return c
\end{minted}
\vspace{.5cm}
\caption{
Our algorithm (top) to determine a particular coefficient $c_n$ of Eq.~\ref{eq:cn_compact}, given an arbitrary complex matrix $G : \mathbb{C}^{2^N \times 2^N}$.
This works by evaluating Eq.~\ref{eq:c_def_in_terms_of_fn}, iterating $i$ in order of the Gray codes, and using the recurrent definition of $\lambda_i^{(n)}$ given in Eq.~\ref{eq:lambda_recurrence_over_i}.
For comparison, we include a slower, alternative algorithm (bottom) which also evaluates Eq.~\ref{eq:c_def_in_terms_of_fn} but wastefully computes $\lambda_i^{(n)}$ in-full for every iteration.
Both schemes permit embarrasingly parallel evaluation of $c_n$ for different $n$.
\\
\\
This implementation is available on Github, at \href{https://github.com/TysonRayJones/DensePauliDecomposer}{\texttt{github.com/TysonRayJones/DensePauliDecomposer}}.
}
\end{subfigure}
\caption{}
\label{fig:alg}
\end{figure*}

On platforms where an unsigned integer (equal to the word size) has at least $2N$ bits, the total runtime to compute a single coefficient $c_n$ scales as $\mathcal{O}(2^N)$, and ergo determining all $4^N$ coefficients takes time $\mathcal{O}(8^N)$. 
When $2N$ exceeds the word size, our method incurs a slowdown of factor $\mathcal{O}(N)$, although this is an unrealistic regime - with modern $64$-bit words, such a scenario would require our input complex matrix $\matr{G}:\mathbb{C}^{2^{32} \times 2^{32}}$, at $8$-bit double precision, was already $256$~\textit{exbibytes} in memory. This is remarkably close to the 2011 estimate of the world's total data storage capacity~\cite{hilbert2011world}.

Ignoring the cost of the input matrix, our scheme uses only a fixed memory overhead in each evaluation of $c_n$. This permits the determination of a $c_n$ from a matrix $\matr{G}$ which may fill almost the entirety of the system's available memory. Serial simulation at such sizes will likely have long become untenable; however, because evaluation of $c_n$ across $n$ is embarrasingly parallel, our scheme is trivial to parallelise.

Indeed, because our scheme does not modify any scaling data structures, the memory overhead of multithreading~\cite{nemirovsky2013multithreading} remains negligible, and we have no risk of performance degradation due to cache conflicts and false sharing~\cite{bolosky1993false}. It is further straightforward to distribute evaluation of $c_n$ across a computer network~\cite{gerla1977topological}. Finally, because our algorithm contains no branching whatsoever, it is well-suited to GPU parallelisation~\cite{han2011reducing}.

\section{Benchmarking}

\begin{figure}[b!]
    \centering
    \includegraphics[width=.7\textwidth]{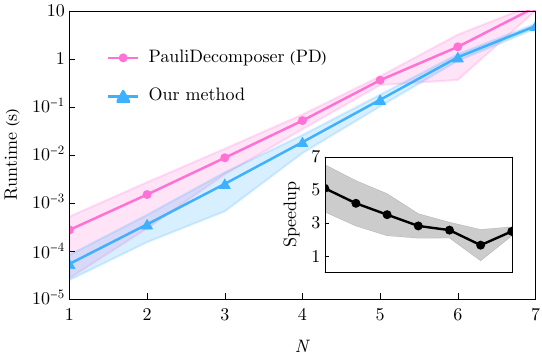}
    \caption{Average runtime (with $3\times$ standard deviations shown in shading) to fully decompose (i.e. calculate all $4^N$ coefficients) one hundred random, dense, complex $2^N\times 2^N$ matrices into $N$-Pauli tensors. The ratio of PD's runtime to that of our method is shown in the subplot.
    }
    \label{fig:benchmarking}
\end{figure}

We compare our method to a very recent technique called the ``PauliDecomposer" (PD) which was demonstrated in Ref.~\cite{romero2023paulicomposer} to outperform several other state-of-the-art methods.
While the PD has an equivalent runtime complexity to our method of $\mathcal{O}(8^N)$, it additionally requires a growing memory overhead of $\mathcal{O}(2^N)$, in contrast to our fixed $\mathcal{O}(1)$ cost.
Note that PD has optimised, asymptotically-improved edge-cases for when the input matrix is diagonal, or entirely real. Such facilities are not implemented by our method, and not investigated by our benchmarking.

We measure the runtime of both algorithms fully decomposing random complex matrices when running serially in Python v3.9.1 on a 2017 13-inch Macbook Pro, with 16\,GiB RAM and a 2.5\,GHz Intel Core i7.
Varying $N$ from $1$ to $7$, we time both algorithms decomposing one hundred random matrices. Our results are shown in Fig.~\ref{fig:benchmarking}, and exhibit a speedup of our method over the PD by a factor $1.5\times$ to $5\times$.

Our largest benchmark is of a meagre $N=7$, whereby the input double-precision matrix $\matr{G}:\mathbb{C}^{2^7 \times 2^7}$ is $256\,$KiB in memory, and the temporary data structures used by the PauliDecomposer are, to the best of the author's assessment, only $2\,$KiB in memory. This is smaller than the tested platform's L1 cache of $32\,$KB. As such, we have not tested the regime where the PauliDecomposer's memory requirements may invoke a runtime penalty due to caching, to which our own constant-memory method is not liable. Benchmarks for larger $N$ might ergo showcase greater speedup, though are presently precluded by the author's laziness.

We note the pseudocode for PD (for \textit{decomposing} a matrix into the Pauli basis, in contrast to the manuscript's other methods for \textit{composing} matrices) is not given in the manuscript's main-text, but is instead available in its accompanying Github repository~\cite{romerogithub}.

\section{Acknowledgements}

We thank Joe Gibbs and Zoë Holmes for helpful discussions, and Manuel Rudolph for unhelpful ones, had between the steam room and the bar during a snowy ill-fated visit to Grindelwald, Switzerland.
This research was supported by the NCCR MARVEL, a National Centre of Competence in Research, funded by the Swiss National Science Foundation (grant number 205602),
and by EPSRC grant EP/M013243/1.
We dedicate our colour scheme to Margot Robbie's \textit{Barbie} (2023) Oscar snub.

\bibliographystyle{unsrt}
\bibliography{bibliography.bib}

\end{document}